\documentclass[prd,superscriptaddress,eqsecnum,nofootinbib,showpacs,preprintnumbers]{revtex4-2}
\usepackage{amsmath}
\usepackage{amssymb}
\usepackage{here}
\usepackage{amsfonts,amsthm,bm}
\usepackage{color,xcolor}
\usepackage[greek,english]{babel}
\usepackage{titlesec}

\usepackage{graphicx,epsfig}
\graphicspath{{figures/}{./}}
\usepackage{float}
\usepackage{subfigure}
\usepackage{tikz}
\newcommand{\be}{\begin{eqnarray}}
	\newcommand{\ee}{\end{eqnarray}}
\newcommand{\bea}{\begin{eqnarray}}
	
	\newcommand{\eea}{\end{eqnarray}}

\def\Ref{\ref}

\newcommand{\beq}{\begin{equation}}
\newcommand{\eeq}{\end{equation}}
\newcommand{\bseq}{\begin{subequations}}
	\newcommand{\eseq}{\end{subequations}}

\usepackage{empheq}

\begin{document}
\preprint[{\leftline{KCL-PH-TH/2022-{\bf 60}}

\title{Global Monopoles in the Extended Gauss-Bonnet Gravity}

\author{Nikos Chatzifotis }
\email{chatzifotisn@gmail.com} \affiliation{Department of Physics, School of Applied Mathematical and Physical Sciences, National Technical University of Athens, 15780 Zografou Campus,
Athens, Greece.}

\author{Nick E. Mavromatos}

 \affiliation{Department of Physics, School of Applied Mathematical and Physical Sciences, National Technical University of Athens, 15780 Zografou Campus,
Athens, Greece.}

\affiliation{Theoretical Particle Physics and Cosmology Group, Department of Physics, King's College London, London WC2R 2LS, UK}

\author{Dionysios P. Theodosopoulos}

\affiliation{Department of Physics, School of Applied Mathematical and Physical Sciences, National Technical University of Athens, 15780 Zografou Campus,
Athens, Greece.}

%\vspace{17.5cm}
%\begin{abstract}

%\end{abstract}
\vspace{1.0cm}

\begin{abstract}
We discuss self-gravitating global O(3) monopole solutions associated with the spontaneous breaking 
of O(3) down to a global O(2) in an extended Gauss Bonnet theory of gravity in (3+1)-dimensions, 
in the presence of a non-trivial scalar field $\Phi$ that couples to the Gauss-Bonnet higher curvature combination with a coupling parameter $\alpha$. We obtain a range of values for $\alpha < 0$ (in our notation and conventions), which are such that a global (Israel type) matching is possible of the space time exterior to the monopole core $\delta$ with a de-Sitter interior, guaranteeing the positivity of the ADM mass of the monopole, which, together with a positive core radius $\delta > 0$, are both dynamically determined as a result of this matching. It should be stressed that in the General Relativity (GR) limit, where $\alpha \to 0$, and $\Phi \to $ constant, such a matching yields a negative ADM monopole mass, which might be related to the stability issues the (Barriola-Vilenkin (BV)) global monopole of GR faces. Thus, our global monopole solution, which shares many features with the BV monopole, such as an asymptotic-space-time deficit angle, of potential phenomenological/cosmological interest, but has, par contrast, a positive ADM mass, has a chance of being a stable configuration, although a detailed stability analysis is pending.

\end{abstract}

\maketitle

\flushbottom

%\tableofcontents

\section{Introduction and motivation: The (self-gravitating) $O(3)$ Global Monopole Solution}\label{sec:intro}

Global monopoles have been suggested for the first time by Barriola and Vilenkin~\cite{vilenkin} (BV)  and are self-gravitating singular objects, which are static solutions to the gravitational and scalar equations of motion of a field theoretical system of a triplet of real scalar fields $\chi^a$, $a=1,2,3$, transforming in the fundamental representation of an O(3) global group, embedded in an Einstein (General-Relativity (GR)) curved space-time.  
The fields have a symmetry breaking potential which breaks spontaneously, through an appropriate vacuum expectation value (v.e.v.) of the triplet, $\eta = {\rm constant} \ne 0$, the O(3) group down to a global O(2). 

The Lagrangian of the BV model is 
\begin{equation}
L=\left(-g\right)^{1/2}\left\{ R -\frac{1}{2}\partial_{\mu}\chi^{a}\partial^{\mu}\chi^{a}-\frac{\lambda}{4}\left(\chi^{a}\chi^{a}-\eta^{2}\right)^{2} \right\} \label{eq:gm}
\end{equation}
where $g_{\mu\nu}$ is the (3+1)-dimensional metric tensor,  $g$ is its determinant  and $R$ is the Ricci scalar for
$g_{\mu\nu}$~\footnote{Our conventions and definitions throughout this work are: $(-,+,+,+)$ for the signature of the metric, the Riemann tensor is defined as 
$R^\lambda_{\,\,\,\,\mu \nu \sigma} = \partial_\nu \, \Gamma^\lambda_{\,\,\mu\sigma} + \Gamma^\rho_{\,\, \mu\sigma} \, \Gamma^\lambda_{\,\, \rho\nu} - (\nu \leftrightarrow \sigma)$, 
and the Ricci tensor and scalar are given by  $R_{\nu\alpha} = R^\lambda_{\,\,\,\,\nu \lambda \alpha}$ and $R= g^{\mu\nu}\, R_{\mu\nu}$ respectively.}.

As a result of the Goldstone theorem, these global monopoles have massless Goldstone fields associated with them. The energy densities of the latter  scale with the radial distance $r$ from the monopole core  as $1/r^2$, which lead to a linear divergence of the monopole total energy density. 
In \cite{vilenkin}, only estimates of the total monopole mass have been given by considering the solution in the exterior of the monopole core. Estimating the core size to be of order $\delta \sim \lambda^{-1/2} \, \eta^{-1}$, 
one obtains a heuristic global monopole particle mass estimate of order 
$M_{\rm core} \sim  \delta^3 \, \lambda \, \eta^4 = \lambda^{-1} \eta $.  
The presence of non-trivial curvature in the portion of spacetime exterior to the monopole core implies that the above estimates should be rethought within the context of the gravitational equations stemming from the Lagrangian \eqref{eq:gm}. However, since the gravitational effects are expected to be weak, given that phenomenologically, in order for the global monopoles to have a chance of being detected, they should have survived inflation, and thus the 
associated O(3) symmetry breaking should occur at scales much lower than the inflationary scale, thus $\eta \ll M_{\rm P}$, with $M_{\rm P}$ the Planck mass, BV argued that the flat space-time estimates for the core mass might still be valid, as an order of magnitude estimate. Nonetheless, as we shall explain below, their naive analysis does not work, given that the back reaction effects of the monopole onto the space-time are significant, and affect its stability. 

Indeed, outside the monopole core, BV used approximate asymptotic analysis of the equations of motion for the scalar fields $\chi^a$, $a=1,2,3$, and the 
Einstein equations, 
\begin{equation}
G_{\mu\nu} \equiv R_{\mu\nu} - \frac{1}{2} g_{\mu\nu} R = 8\pi G_{\rm N} \, T^\chi_{\mu\nu}
\end{equation}
with $G_{\mu\nu}$ the Einstein tensor, and $T_{\mu\nu}^\chi$ the matter stress tensor derived from the Lagrangian (\ref{eq:gm}). The scalar field configuration for a global monopole is~\cite{vilenkin}
\begin{equation}\label{eq:gmf}
\chi^a = \eta \, h(r) \, \frac{x^a}{r}~, \quad a=1,2,3 
\end{equation}
where $x^a$ are spatial Cartesian coordinates, $r = \sqrt{x^a x^a }$ is the radial distance, and $h(r) \to 1 $ for  $r \gg \delta $. So 
at such large distances, the amplitude squared of the scalar field triplet approaches the square of the vacuum expectation value $\eta$, $\chi^a \chi^a \to \eta^2$. 

As a result of the symmetry breaking, the space-time, for $r \gg \delta $, differs from the standard Schwarzschild metric corresponding to a massive object with mass $M_{\rm core}$ (assuming that all the mass of the monopole is concentrated in the core's interior):
\begin{equation}\label{asympt}
ds^2  = - \Big( 1 - 8\pi \, G_{\rm N} \eta^2 - \frac{2 G_{\rm N}\, M_{\rm core}}{r} \Big) dt^2 + \frac{dr^2}{1 - 8\pi \, G_{\rm N}\, \eta^2 + \frac{2 G_{\rm N}\, M_{\rm core}}{r}} + r^2 \Big( d\theta^2 + {\rm sin}^2 \theta \, d\phi ^2 \Big)~, \quad r \gg \delta~,
\end{equation}
where $G_{\rm N}$ is the (3+1)-dimensional Newton constant and 
$(r,\theta,\phi)$ are spherical polar coordinates. As a result of the symmetry-breaking $\eta$ v.e.v., the
space-time is not asymptotically ($r \to \infty$) flat, but differs from  Minkowski by the presence of a conical \emph{deficit solid angle} $\Delta \Omega = 8\pi G_{\rm N} \, \eta^2$. \color{black} The Schwarzschild metric is obtained in the unbroken phase ($\eta \to 0$). The space-time (\ref{asympt}) becomes a Minkowski space-time with 
\begin{equation}\label{asymptflat}
ds^2  = - d{t^\prime}^2 + d{r^\prime}^2 +   \Big( 1 - 8\pi \, G_{\rm N} \eta^2 \Big)\, {r^\prime}^2 \Big( d\theta^2 + {\rm sin}^2 \theta \, d\phi ^2 \Big)~, \quad r \gg \delta~,
\end{equation}
where in arriving at \eqref{asymptflat} we have rescaled the time $t \to (1 - 8\pi \, G_{\rm N} \, \eta^2)^{-1/2} \, t^\prime $, and radial coordinate $r$, $r \to (1 - 8\pi \, G_{\rm N} \, \eta^2)^{1/2}\, r^\prime$. 
We stress that the space-time (\ref{asymptflat}) is not flat, since the scalar curvature behaves 
\color{black}  
\begin{align}\label{scalardef}
R \propto \frac{16\,\pi \,G_{\rm N} \, \eta^2}{r^2}\,, \quad r \to \infty\,.
\end{align}
\color{black} 
The presence of such a monopole-induced deficit solid angle, can have important physical consequences for scattering processes in such space-times: the scattering amplitude in the forward direction is very large~\cite{papavassiliou} in 
angular regions of order of the deficit angle (or equivalently the squared ratio of the monopole mass to the Planck mass). Thus one may have interesting methods~\cite{papavassiliou} for looking for global monopoles in, e.g., the early Universe, where the scattering of cosmic-microwave-background photons off them, can lead to Einstein rings of strong intensities that could constitute characteristic patterns.

However, the work of \cite{vilenkin}, has sparked a still ongoing debate regarding the stability of the configuration~\cite{debate}. Contributing formally to this debate, supporting arguments in favour of instability of the solution of the global monopole, is the detailed analysis of the gravitational back reaction effects of such defects onto the space time, performed in \cite{negative}. In that work, the back reaction has been determined by requiring a matching of the solutions of the non-linear coupled system of gravitational and matter equations at the core radius, by replacing the interior to the core radius space-time by a de-Sitter one, with positive cosmological
constant. The singularity at $r \to 0$ is {\it regularised}, and  
the core size is determined dynamically rather than heuristically from flat space arguments as was done in \cite{vilenkin}.  The results of  \cite{negative} amounts to a core radius 
\begin{align}\label{core}
r_c = 2 \, \lambda^{-1/2}\, \eta^{-1}\,, 
\end{align}
for the self-gravitating solution, which is found by matching an exterior  Schwarzschild-like metric 
$$ ds^2 = - \Big(1 - 8\pi G_{\rm N}\, \eta^2 \, - \frac{2\, G_{\rm N}\, M}{r} \Big) dt^2 + \Big(1 - 8\pi G_{\rm N}\, \eta^2 \, - \frac{2\, G_{\rm N}\, M}{r} \Big)^{-1}
 dr^2 + r^2 \, d\Omega^2~, $$ to an interior local de Sitter metric 
$$ ds^2 = - \Big(1 - \mathcal{H}^2 \, r^2 \Big) dt^2 + \Big(1 - \mathcal{H}^2 \, r^2 \Big)^{-1} dr^2 + r^2 \, d\Omega^2~ $$
where $M$ denotes the monopole mass and $\mathcal{H}^2 = \frac{8\pi G_{\rm N}\, \lambda \, \eta^4}{12}$ the de-Sitter parameter. The motivation to use such a matching comes from the observation that at the origin ($r \to 0$) the Higgs potential for the scalars leads to a cosmological constant $\propto \eta^4$, since any ``matter'' scalar  fields go to zero. Thus one does not have any other structures in the interior of the core, except perhaps a cosmological constant, which could be a vacuum effect. 
Unfortunately, however, such a matching yields a negative mass for the monopole, 
\begin{align}\label{negmass}
M\sim -6\pi \lambda^{-1/2} \eta <0\,,
\end{align}
which may be viewed as supporting the instability of the BV global monopole. 

In~\cite{negative} such a negative mass has been interpreted as a consequence of the repulsive nature of gravity induced by the vacuum-energy $H^2$ provided by the global monopole. A classification of the space-times arising from a self-gravitating global monopole solution of the type considered in \cite{vilenkin} and in \cite{negative}, \emph{i.e.} in field theories with only the triplet of the Higgs-type scalar fields and the Ricci scalar curvature, has been given in \cite{bronnikov}, where it was argued that, upon requiring \emph{regularity at the centre} of the monopole, but otherwise independently of the shape of the Higgs potential, the metric can contain at most one horizon. In such a case with a single horizon, the global space-time structure is that of a de-Sitter space-time.

However, if a black hole or other geometric singularity is present as $r \to 0$, like in the string-inspired 
model of \cite{sarkarNEM}, in which the self-gravitating BV model  \eqref{eq:gm} is embedded in the framework of the effective action originating from the bosonic strings, involving antisymmetric tensor and electromagnetic fields, which induce a Reissner-Nordstrom geometry in the interior of the core of the initial BV configuration, 
then the space-time for small $r$ ($r \to 0$)  is different. As shown in \cite{sarkarNEM}, then, the argument leading to negative mass would not  hold, and one can obtain a positive mass global monopole in such 
extended string-inspired gravity theories. 

The presence of electromagnetic fields in this string-inspired model, though, elevates the global monopoles to magnetic monopoles, as it induces a singular-like magnetic field $\mathcal B \sim 1/r^2$, of the type encountered in the standard case of a magnetic monopole~\cite{shnir,mitsou}. 
It should be stressed for completeness that in the model of \cite{sarkarNEM}, it is the equation of motion of the dilaton that connects the antisymmetric Kalb-Ramond parts of the effective action to the electromagnetic sector.
The KR tensor in (3+1)-dimensions is dual to a massless axion-like particle. In this case, the magnetic charge is proportional to the axion ``charge'', which is a constant $\zeta$ that characterises the $ \zeta/r$ asymptotic ($r \to \infty$) behaviour  of the axion field.

After the above discussion, a natural question arises whether one can find global monopoles \`a l\`a BV in extended gravity models, in which the de-Sitter regularisation of \cite{negative} leads to positive mass, as in the model of \cite{sarkarNEM}, but here within the gravitational sector only, without elevating the global monopole to a magnetic one. This question will be answered in the affirmative in this article, where we shall show that a global-monopole solution of the form of \cite{vilenkin}, embedded in the higher curvature gravitational theory of ~\cite{Fernandes:2021dsb}, which we call from  now on extended Gauss Bonnet (eGB), can undergo the de-Sitter regularization of the singularity as in \cite{negative}, maintaining a {\it positive} finite mass. This would imply that, contrary to the standard BV global monopoles of \cite{vilenkin}, such global monopoles might be stable, behaving as ordinary ``particle'' excitations. 
We stress here, though, that the positivity of the mass of the regularised configuration does not constitute a proof of stability. Such a task will not be the topic of the current discussion.

The structure of the article is the following: in section \ref{sec:eGB} we review the basic features of the eGB gravitational theory, whilst \color{black} in \ref{sec:gmeGB} we discuss the analytic parts of the global monopole solution embedded in eGB theory, in the asymptotic regions near the origin ($r=0$), the core region ($r \sim \delta$) and at infinity ($r \to \infty$). We study 
the de-Sitter regularisation of the $r \to 0$ singularity, and the resulting positivity of the associated mass. In section \ref{sec:numer} we give the numerical solution interpolating between the aforementioned asymptotic regions of space, 
and we compare with our analytic treatment, which completes the discussion on the existence of such solutions to the eGB theory. \color{black} Conclusions and outlook are given in \ref{sec:concl}.

\section{Extended Gauss-Bonnet Gravity}\label{sec:eGB}

 In this section we review the eGB gravitational theory, as presented in \cite{Fernandes:2021dsb}, which will be our framework for the derivation of global monopoles. The theory has the advantage of containing a scalar field degree of freedom, which is conformally coupled. This yields a purely gravitational field equation, which may act as a constraint on the equations of motion, thus allowing for closed-form local solutions to be found. In general, the complexity of scalar-tensor theories seldomly allows for exact analytical local solutions. However, by imposing additional symmetries in the underlying gravitational action, exact solutions may be derived, which is exactly the case here. The action of the theory reads
 \begin{equation}
 	\label{action}
 	S_{\rm eGB}=\int \frac{d^4 x \sqrt{|g|}}{16\pi G_{\rm N}}\left[R-\beta e^{2\Phi}(R+6(\nabla\Phi)^2)-2\lambda e^{4\Phi}-\alpha(\Phi\, \mathcal{G}-4 G_{\mu\nu}\nabla^\mu\Phi\nabla^\nu\Phi-4\square\Phi (\nabla\Phi)^2-2(\nabla\Phi)^4)\right]
 \end{equation}
where $\mathcal{G}=R_{\alpha\beta\mu\nu}R^{\alpha\beta\mu\nu}-4R_{\mu\nu}R^{\mu\nu}+R^2$ is the Gauss-Bonnet term. It should be noted that in the limit of vanishing coupling constants $\alpha$ and $\lambda$ one obtains the usual Einstein gravity with a conformally coupled scalar field \cite{Bocharova:1970skc,Bekenstein:1974sf}. In a sense, the idea of a conformally coupled scalar field in extended gravitational theories is certainly not new. The advantage of the action under consideration is that it allows for hairy black hole solutions with the scalar field being everywhere regular but the origin. It should also be stressed that the main form of the theory was previously derived by Lu and Pang in their pioneering paper \cite{Lu:2020iav}, where they considered the singular dimensional limit of the D-dimensional Lovelock action. This establishes a beautiful connection of the compactified D-dimensional Lovelock gravity to the most general scalar-tensor gravity with a conformally coupled scalar field \cite{Fernandes:2022zrq}.\\
	
	Variation of the action \eqref{action} with respect to the metric yields the following equations of motion
	\begin{equation}
		\label{graveom}
	(\mathcal E_{\mu\nu}): \quad 	G_{\mu\nu}=-\alpha \mathcal{H}_{\mu\nu}+\beta e^{2\Phi} \mathcal{A}_{\mu\nu}-\lambda e^{4\Phi}g_{\mu\nu}
	\end{equation}
	where	
	\begin{equation}
		\label{GBterm}
		\begin{split}
		\mathcal{H}_{\mu\nu}=2G_{\mu\nu}(\nabla\Phi)^2+4P_{\mu\alpha\nu\beta}(\nabla^\alpha\Phi\nabla^\beta\Phi-\nabla^\alpha\nabla^\beta\Phi)+4(\nabla_\alpha\Phi\nabla_\mu\Phi-\nabla_\alpha\nabla_\mu\Phi)(\nabla^\alpha\Phi\nabla_\nu\Phi-\nabla^\alpha\nabla_\nu\Phi)\\
		+4(\nabla_\mu\Phi\nabla_\nu\Phi-\nabla_\mu\nabla_\nu\Phi)\square\Phi+g_{\mu\nu}\left(2(\square\Phi)^2-(\nabla\Phi)^4+2\nabla_\alpha\nabla_\beta\Phi(2\nabla^\alpha\Phi\nabla^\beta\Phi-\nabla^\alpha\nabla\beta\Phi)\right),
			\end{split}
	\end{equation}
	and 
	\begin{equation}
		\label{ricciterm}
		\mathcal{A}_{\mu\nu}=G_{\mu\nu}+2\nabla_\mu\Phi\nabla_\nu\Phi-2\nabla_\mu\nabla_\nu\Phi+g_{\mu\nu}\left(2\square\Phi+(\nabla\Phi)^2\right)
	\end{equation}
with $\displaystyle P_{\alpha\beta\gamma\delta}=-\frac{1}{4}\epsilon_{\alpha\beta\mu\nu}R^{\mu\nu\kappa\lambda}\epsilon_{\kappa\lambda\gamma\delta}$ denoting the double dual of the Riemann tensor and  $\epsilon_{\alpha\beta\gamma\delta}$ the Levi-Civita tensor with the usual convention of $\epsilon_{0123}=\sqrt{|g|}$. On the other hand, variation of the action \eqref{action} with respect to the scalar field $\Phi$ yields the equation
 \begin{equation}
 	\label{scalar}
 	\beta\tilde{R}+\frac{\alpha}{2}\tilde{\mathcal{G}}+4\lambda=0
 \end{equation}
where $\tilde{R}$ and $\tilde{\mathcal{G}}$ are the Ricci scalar and the Gauss-Bonnet term computed on the transformed metric $\tilde{g}_{\mu\nu}=e^{2\Phi}g_{\mu\nu}$. By virtue now of the scalar field being conformally coupled, one may deduce that the combination of the equations of motion $\displaystyle -2g_{\mu\nu}\frac{\delta S_{\rm eGB}}{\delta g_{\mu\nu}}+\frac{\delta S_{\rm eGB}}{\delta \Phi} = 0$ yields the condition :
\begin{equation}
	\label{conformal}
	R+\frac{\alpha}{2}\mathcal{G}=0\,,
\end{equation}
which establishes a helpful geometric constraint for the derivation of local solutions. In particular, due to (\Ref{conformal}), one may set a spherically symmetric homogeneous metric ansatz of the form
\begin{equation}
	\label{ansatz}
	ds^2=-f(r)dt^2+\frac{dr^2}{f(r)}+r^2d\Omega^2
\end{equation}
and verify the metric component solution of {\color{black}\cite{Fernandes:2021dsb} given as}
\begin{equation}
	\label{metric}
	f(r)=1+\frac{r^2}{2\alpha}\left[1-\sqrt{1+4\alpha\left(\frac{2 G_{\rm N} M}{r^3}+\frac{C}{r^4}\right)}\right]
\end{equation}
where $C$ is an integration constant to be verified by the equations of motion, while $M$ is the ADM mass of the spacetime. Following on this result, the scalar field solution for this metric may be derived by the combination of the gravitational equations  $\mathcal{E}^{t}_t-\mathcal{E}^r_r=0$ ({\it cf.} \eqref{graveom}), which allows for the following non-trivial scalar field profiles to be found:
\begin{align}
	\label{scalar1}
	\Phi (r)&=\ln \left(\frac{c_1}{r+c_2}\right)\\
	\label{scalar2}
	\Phi (r)&=\ln \left(\frac{\sqrt{\frac{-2\alpha}{\beta}}}{r} {\rm sech}\left(c_3\pm\int^r \frac{dr^\prime}{r^\prime\sqrt{f(r^
	\prime)}}\right)\right)
\end{align}
The integration constants for the metric and the scalar profiles are to be determined by the rest of the equations at hand. Interestingly, each of the scalar field profiles imposes a different integration constant on the metric component and a different value of the $\lambda$ parameter. In particular, for the first case of (\Ref{scalar1}), one finds that $C=2\alpha$, $c_1=\sqrt{-2\alpha/\beta}$ and $c_2=0$, while $\displaystyle \lambda=\frac{\beta^2}{4\alpha}$. For the second case of (\Ref{scalar2}), one finds that $C=0$ with $\displaystyle \lambda=\frac{3\beta^2}{4\alpha}$, while $c_3$ is left undetermined. Moreover, both solutions imply that the coupling constants $\alpha$ and $\beta$ need to satisfy $\alpha\beta<0$. An overview of the action (\Ref{action}) shows that if $\beta<0$, then the canonical kinetic term of the scalar field has the wrong effective sign, which implies a scalar field of phantom nature. To this end, we constrain ourselves to $\beta>0$ and $\alpha<0$, as was the case in the analysis of the corresponding black-hole solutions performed in \cite{Babichev:2022awg}.

\section{Global Monopoles with Positive Mass  in Extended Gauss-Bonnet Gravity}\label{sec:gmeGB}

Having paid our dues to the review of the gravitational theory and its corresponding local solutions, we may now move on to the main topic of our work, which is the derivation of global monopole spacetimes. In particular, we will show that the negative $\alpha$ parameter appearing in the action may stabilize the global monopole spacetime, which in GR has the known pathology of yielding negative mass \cite{vilenkin, negative}. To extract monopole spacetimes, we consider the addition of a Higgs triplet matter content $\chi^a$, $a=1,2,3$, and its corresponding Higgs potential {\color{black}$\displaystyle V(\chi)=\frac{\xi}{4}(\chi^a\,\chi^a-\eta^2)^2$} on the gravitational theory,
\begin{equation}\label{scal}
	S=S_{\rm eGB}+\int d^4x \sqrt{|g|}\left[-\frac{1}{2}(\partial_\mu\chi^a)(\partial^\mu\chi^a)-\frac{\xi}{4}(\chi^a\,\chi^a-\eta^2)^2\right],
\end{equation}
which in the unbroken sector satisfies an internal O(3) symmetry. The corresponding gravitational equations of motion are trivially modified with the addition of the Higgs stress-energy tensor {\color{black}$\displaystyle T_{\mu\nu}^{\chi}=\nabla_\mu \chi^a\nabla_\nu \chi^a-\frac{1}{2}g_{\mu\nu}(\nabla\chi)^2-g_{\mu\nu}V(\chi)$}, i.e.  
\begin{equation}
	\label{graveomhiggs}
	G_{\mu\nu}=-\alpha \mathcal{H}_{\mu\nu}+\beta e^{2\Phi} \mathcal{A}_{\mu\nu}-\lambda e^{4\Phi}g_{\mu\nu}+8\pi \,G_{\rm N}\, T_{\mu\nu}^{\chi}
\end{equation}
while the Higgs equation of motion is simply
\begin{equation}
	\label{higgs}
	\square\chi^a=\xi\chi^a (\chi^b \chi^b - \eta^2)\, \quad a,b=1,2,3\,.
\end{equation}
On the other hand, the previous geometric condition of (\Ref{conformal}) will also contain the Higgs matter content and reads
\begin{equation}
	\label{conformalhiggs}
	R+\frac{\alpha}{2}\mathcal{G}+8\pi \, G_{\rm N} \, T^{\chi}=0,
\end{equation}
where $T^{\chi}=g^{\mu\nu}T_{\mu\nu}^{\chi}$. In order for the action to contain a global monopole solution, one uses the familiar ansatz for the Higgs field of ({\it cf.} \eqref{eq:gmf},  section \ref{sec:intro}):
\begin{equation}
	\label{hansatz}
	\chi^a = \eta \,  h(r)\frac{x^a}{r},\quad x^a x^a=r^2
\end{equation}
with $h(r)\rightarrow 1$ as $r\rightarrow \infty$. Note that this constraint also implies that $\partial_\mu h(r)\rightarrow 0$ as $r\rightarrow \infty$. Under (\Ref{hansatz}), one may write the components of the Higgs triplet as
\begin{equation}
	\label{isospace}
	x^1=r \cos\phi \sin\theta, \quad x^2=r\sin\phi \sin\theta, \quad x^3=r\cos\theta
\end{equation}
and the stress-energy tensor may be expressed as a tensor dependent solely on the $h(r)$ function, while the equation of motion for the Higgs triplet is reduced to a single differential equation of $h(r)$. In particular, keeping the homogeneous spherically symmetric ansatz of (\Ref{ansatz}), we may find that (\Ref{higgs}) is reduced to
\begin{equation}
	\label{higgseom}
	\left(h''+\frac{2h'}{r}\right)f+h' f'-\xi \eta^2 h^3+\left(\xi\eta^2-\frac{2}{r^2}\right)h=0\,,
\end{equation}
\color{black} where the prime denotes derivativce with respect to $r$.\color{black}

In the presence of the Higgs sector, the equations of motion are to be solved asymptotically in the regions of large $r$ and $r\rightarrow 0$.
We will firstly focus on the asymptotic region, where we may set $h(r)\rightarrow 1$. We note that this approximation satisfies the Higgs equation of motion only up to $\mathcal{O}(r^{-1})$.
\color{black} We note at this stage that, upon setting $h(r) = 1 + \mathcal O(r^{-1})$ in \eqref{higgseom}, for asymptotically  large $r \to \infty$, this equation does not yield any information on the function $f(r)$ to this order in $1/r$. 
To obtain the asymptotic solution for $f(r)$, $r \to \infty$, we plug the approximation of a constant Higgs, $h(r)\approx 1$, into the condition of (\Ref{conformalhiggs}), and find that \color{black}
\begin{equation}
	\label{metrichiggs}
		f(r)=1+\frac{r^2}{2\alpha}\left[1-\sqrt{1+4\alpha\left(\frac{2 G_{\rm N} M}{r^3}+\frac{C}{r^4}+\frac{\eta^2 8\pi G_{\rm N}}{r^2}\right)}\right]
\end{equation}
which implies that the monopole contribution yields a solid deficit angle on the underlying spacetime, which is a familiar result from the global monopole solution of GR of, \cite{vilenkin}, reviewed in section \ref{sec:intro}. Indeed, it is easily verified {\color{black}from the solution of (\Ref{metrichiggs})} that 
\begin{equation}
	\label{asym}
	f(r)\sim 1-\eta^2 \, 8\pi \, G_{\rm N} - \frac{2 G_{\rm N} M}{r}+O(r^{-2}),\qquad r\rightarrow \infty\,.
\end{equation} 
\color{black} The reader is reminded of the non trivial scalar curvature \eqref{scalardef} of order $\mathcal O(r^{-2})$, of an asymptotic ($r \to \infty$) spacetime with a deficit, of the form \eqref{asym}, which is a characteristic feature of such constructions, characterised by a broken symmetry~\cite{vilenkin}. \color{black}

It should be noted that in the unbroken phase where $\eta=0$, we recover the original local solution of the theory. However, the choice of the scalar profile will play a crucial role in stabilizing the monopole spacetime. We note that, since the scalar profile equation is derived from $\mathcal{E}^t_t-\mathcal{E}^r_r$ ({\it cf.} \eqref{graveom}) and the Higgs sector contribution vanishes in this combination for the asymptotic value of $h(r)\rightarrow 1$, the scalar field solutions will be the same as (\Ref{scalar1}) and (\Ref{scalar2}). We shall choose the scalar field profile (\Ref{scalar1}), since this is the only one that can stabilize the monopole spacetime, as we will later on comment on the results. We note that this solution again yields the following values for the corresponding integration constants:
\begin{equation}
	\label{const}
	C=2\alpha,\quad c_1=\sqrt{-2\alpha/\beta},\quad c_2=0
\end{equation} 
while the $\lambda$ parameter is \color{black} fixed at 
\begin{align}\label{lambeta}
\displaystyle \lambda=\frac{\beta^2}{4\alpha}\,,
\end{align}
in order to have a smooth limit to the black hole solution with $\eta\rightarrow 0$. \color{black} \\

Moving on to the small $r$ limit, we are focusing on the interior of the monopole. In this region, it is natural to consider that {\color{black} the Higgs field vanishes, while the scalar field $\Phi$ approaches some constant value.\color{black}} This implies that we need to set that
\begin{equation}
	\label{int}
	h(r)\rightarrow0,\quad \Phi(r)\rightarrow \Phi_0,\qquad\text{for}\quad r\rightarrow 0
\end{equation}
while keeping the fixed value of $\displaystyle \lambda=\frac{\beta^2}{4\alpha}$. For this case, the metric component resembles a de-Sitter core for the interior spacetime, {\it i.e.}
\begin{equation}
	\label{intmetric}
	f(r)=1+\frac{r^2}{2\alpha}\left(1-\sqrt{1+\frac{\alpha\, \xi \, \eta^4 \, 8\pi \, G_{\rm N}}{3}}\right)
\end{equation}
as can be verified from (\Ref{conformalhiggs}). We note that this solution constrains the possible values of the $\alpha$ parameter to
\begin{equation}
	\label{llimit}
	\alpha\geq\frac{-3}{\xi\eta^4 \, 8\pi \, G_{\rm N}}
\end{equation}
This range of values yields that (\Ref{intmetric}) indeed describes a de-Sitter core, since the term in the parenthesis is positive definite, while $\alpha<0$. This is the first verification that we correctly chose $\alpha<0$ for our gravitational action. However, this configuration does not exactly solve the equations of motion. {\color{black} One needs also to fix the value of the dilaton in the near origin regime. In order for the equations of motion to be satisfied in this configuration, we find that  
\begin{equation}
	\label{dilaton}
	\Phi_0=\frac{1}{2}\ln\left[\frac{3}{\beta }-\frac{\sqrt{9-3 \eta ^4 \kappa ^2 \xi  \left| \alpha \right| }}{\beta }\pm\frac{\sqrt{\beta ^2 \eta ^4 \kappa ^2 \xi  \left| \alpha \right| +\beta ^2 \left(9-3 \eta ^4 \kappa ^2 \xi  \left| \alpha \right| \right)-4 \beta ^2 \sqrt{9-3 \eta ^4 \kappa ^2 \xi  \left| \alpha \right| }+3 \beta ^2}}{\beta ^2}\right]\,,
\end{equation}
which can be verified that it is a well defined value, i.e. $\Phi_0\in \mathbb{R}$, as long as (\Ref{llimit}) holds. Note that $\beta$ is a free (positive) parameter  and cannot be constrained in the near origin regime. } \color{black} In order to obtain a more complete treatment, a numerical analysis of the equations of motion needs to be implemented, which is 
discussed in the next section \ref{sec:numer}. We remark, though, that, as becomes evident from our analytic treatment in this work and matching using Israel conditions at the monopole core and asymptotically for large $r \to \infty$, the existence of a numerical curve for a solution joining these two regimes smoothly is expected, as we indeed verify explicitly in section \ref{sec:numer}. \color{black}\\

Since the metric components of (\Ref{metrichiggs}) and (\Ref{intmetric}) describe the same spacetime in different patches of the manifold, they need to be matched at an intermediate radius in order to obtain a complete picture of the spacetime. This will yield us the possible values of the ADM mass of the monopole. \\

In order to perform the matching, we follow the Israel conditions, where we match the interior metric component of (\Ref{intmetric}) to the exterior of (\Ref{metrichiggs}), while equating their first normal derivatives on the intermediate radial value to avoid discontinuities on the metric tensor. In particular, we are considering a global metric of the form
\begin{equation}
	\label{metricglobal}
	ds_{G}^2=-F(r)dt^2+\frac{dr^2}{F(r)}+r^2d\Omega^2.
\end{equation}
 $F(r)$ is a distribution function defined as
 \begin{equation}
 	\label{matchmetric}
 	F(r)=F_1(r)\Theta(\delta-r)+F_2(r)\Theta(r-\delta)
 \end{equation}
where $\displaystyle F_1(r)=1+\frac{r^2}{2\alpha}\left(1-\sqrt{1+\frac{\alpha \xi  \eta^4 8\pi G_{\rm N}}{3}}\right)$, 
$\displaystyle F_2(r)=1+\frac{r^2}{2\alpha}\left[1-\sqrt{1+4\alpha\left(\frac{2 G_{\rm N} M}{r^3}+\frac{2\alpha}{r^4}+\frac{\eta^2 8\pi G_{\rm N}}{r^2}\right)}\right]$,  and $\Theta(r)$ denotes the Heaviside function, while $\delta$ is the intermediate radial value where the matching is assumed to be performed. Then, the Israel conditions simply read
\begin{equation}
	\label{israel}
	F_1(r)|_{r=\delta}=F_2(r)|_{r=\delta},\qquad \frac{d}{dr}F_1(r)|_{r=\delta}=\frac{d}{dr}F_2(r)|_{r=\delta}\,.
\end{equation}
Solving the above system with respect to $\delta$ and the ADM mass, we find four solutions, only one of which yields positive values for both quantities. In particular, 
\begin{align}\label{deltaM}
	\delta&=\sqrt{2}\sqrt{\frac{\kappa+\sqrt{\kappa^2-2\alpha\xi}}{\xi\eta^2\kappa}}, \nonumber \\
	M&=\frac{\sqrt{2}\eta}{3 \,G_{\rm N}}(-2\kappa+\sqrt{\kappa^2-2\alpha\xi})\sqrt{\kappa\frac{\kappa+\sqrt{\kappa^2-2\alpha\xi}}{\xi}}
\end{align}
where $\kappa^2=8\pi G_{\rm N}$. Note that the ADM mass is positive if the $\alpha$ parameter is constrained from above as $\displaystyle \alpha<-\frac{3\kappa^2}{2\xi}$, which yields the effective range of 
\begin{equation}
	\label{rangealpha}
	\frac{-3}{\xi\eta^4 \kappa^2}\leq	\alpha <-\frac{3\kappa^2}{2\xi}.
\end{equation}
The existence of an upper limit of negative $\alpha$ was to be expected from the global monopole results of GR~\cite{negative}. Indeed, if $\alpha$ was allowed to reach $\alpha\rightarrow 0$, the ADM mass of the monopole would yield
\begin{equation}\label{limM}
	M_{\alpha\rightarrow0}=-\frac{16\pi \eta}{3\sqrt{\xi}}\,,
\end{equation}
that is, the familiar result of the negative mass of the global monopole we know from GR~\cite{negative}.\footnote{We note at this stage, that in this respect, the four-dimensional Gauss-Bonnet-gravity-$\Phi$ model
of \cite{Lu:2020iav}, corresponds to positive $\alpha >0$, and hence such a model does not include global monopoles with positive ADM mass.} It should be noted here that, having chosen the second scalar profile, the result of the matching is exactly the same as GR and in this case, the parameter $\alpha$ plays no role in the derivation of the monopole mass. This is because we are effectively matching the square root of the interior and the exterior metric and it is the term $2\alpha/r^4$ in the exterior metric which drives the stabilization of the monopole. For the scalar field profile of (\Ref{scalar2}), this term is absent. Our main result, therefore, is that in the presence of the non-trivial extended Gauss-Bonnet term, the repulsive gravitational nature of the global monopole is remedied. 

\section{Numerical Solution}\label{sec:numer}

\color{black} A numerical analysis of the standard global BV monopole regularised with a de Sitter space time in its core has been performed in \cite{negative}, using the Runge-Kutta routine. In our case, as we shall see, the presence of the dilaton and its rapidly changing form (compared to the other fields in the problem (metric and Higgs)) complicates the situation, and forces us to use other methods to construct the numerical solution. 

In what follows we discuss first the validity of the metric function asymptotic solutions, equations (\ref{metrichiggs}) and (\ref{intmetric}), and the continuation of the Higgs $h(r)$ and metric $f(r)$ functions  in the region close to the monopole core ($r=\delta$). In particular, in the spirit of \cite{negative}, we will present the numerical method followed, which leads to the curves shown in figures \ref{fig:1} and \ref{fig:2}. On assuming a more general metric than \eqref{metricglobal}, of the form
\begin{equation}
	\label{ansatz2}
	ds^2=-G(r)dt^2+\frac{dr^2}{f(r)}+r^2d\Omega^2\,,
\end{equation}
we solve the resulting Einstein equation $\mathcal{E}^r_r$ with respect to $G'(r)/G(r)$. We do not present the explicit expressions here, due to their algebraically awkward form, which in any case is not very illuminating. It is important to stress, however, that the ansatz (\ref{ansatz2}) is indispensable, since we have no guarantee that the metric is homogeneous in the non-asymptotic regions. To obtain the numerical solution, we first eliminate $G(r)$ from equations (\ref{conformalhiggs}) and (\ref{higgseom}). Subsequently, we solve these equations along with equation $\mathcal{E}^t_t$ with respect to $h''(r)$, $f'(r)$ and $\Phi''(r)$. Thus, we have three differential equations to be numerically solved simultaneously. 

This problem seems to be stiff, i.e. certain numerical methods for solving the aforementioned equations are numerically unstable, unless the step size is taken to be extremely small. We remind the reader that, an ordinary differential equation problem is stiff if the solution being sought is varying slowly, but there are nearby solutions that vary rapidly, so the numerical method must take small steps to obtain satisfactory results. This is exactly the case with the scalar field $\Phi(r)$ in our problem. Therefore, we solve the differential problem exploiting the "StiffnessSwitching" method of Mathematica, which uses a pair of extrapolation methods as the default. To be more precise, the stiff solver uses the Linearly Implicit Euler method, while the non-stiff solver uses the Explicit Modified Midpoint method. To avoid infinite expressions, we consider initial conditions for $r=\epsilon\approx \mathcal{O}(10^{-5})$: 
\begin{align}\label{bcvalues}
f(\epsilon)\approx 1\,, \, h(\epsilon)\approx 0.5\cdot \epsilon\,, \, h'(\epsilon)\approx 0.5\,, \, \Phi(\epsilon)\approx 4.47\,, \, {\rm and} \, \, \Phi'(\epsilon)\approx 0.001\,,
\end{align}
This particular choice of initial conditions was a challenging problem, which will be discussed later on. Also, we fix the values of the parameters of our model to the set: 
 \begin{align}\label{paramset}
 \eta=0.01\,, \, \xi=10^{4}\, , \, \alpha=-1.6\cdot 10^{-4}\,{\rm and} \, \, \beta=8\cdot 10^{-9}. 
 \end{align}

Moreover, we should recall  that, in order for the correct asymptotic solutions to be obtained, one must have \eqref{lambeta}, with $\Phi_{0}$ fixed by equation (\ref{dilaton}). We work in units $\kappa=1$. For the chosen set of parameters \eqref{paramset}, the initial value of the scalar field reads $\Phi_{0}=0$. Note that the above set of parameters satisfies the inequality (\ref{rangealpha}), and hence, we work with a structure with positive ADM mass. 

Before proceeding further, it is crucial to discuss the above choices of the values of the parameters and the initial conditions. We chose $\sqrt{\xi}\eta=1$ and a value for $\alpha$ close to its upper limit $\alpha_{max}=-1.5\cdot 10^{-4}$ in order to compare our results with those of the numerical analysis of the BV model in \cite{negative}. For instance, for the chosen set of parameters, the radius of the monopole core in our model is $\delta \approx 2.47$, while in BV model reads $\delta_{BV}=2$. Also, we chose the initial conditions of the Higgs function to be similar to those presented in \cite{negative}, in order to obtain $h(r\rightarrow\infty)\rightarrow 1$. The most challenging aspect of our analysis is the choice of $\Phi(\epsilon)$. In order for the metric function to have a plateau $f(r\rightarrow\infty)\rightarrow 1-\eta^{2}=0.9999$, we fix $\Phi(\epsilon)\approx 4.47$. We note at this point that, if we had chosen $\Phi(\epsilon)\approx\Phi_{0}=0$, then the metric function would have asymptotically reached the value $0.99994$, which is not the correct boundary condition of our solution. Hence, we conclude that, close to the origin, the value of the scalar field surged from $0$ to approximately $4.47$, which means that the scalar field acts like a step function near $r=0$. Additionally, on account of our theoretical analysis in section III, we expect that the scalar field is approximately constant inside the monopole core of radius $\delta=2.47$ up to $r\simeq \mathcal{O}(\epsilon)$. This is exactly the case, since we need to fix $\Phi'(\epsilon)=0.001$ in order for a standard form of the Higgs function \cite{negative} and a reasonable form of the metric function to be obtained.

Around the monopole core, the numerical problem becomes extremely difficult to handle, because the derivative of the scalar field around the core, $\Phi'(r\simeq\delta)$, is rapidly changing.
This implies that either 
more advanced numerical methods are required or an interpolation method to match the interior with the exterior solutions should be implemented. We chose the second method, where a numerical analysis has been applied in regions $r\in (\epsilon,\delta-0.1)$ and $r\in (\delta+0.1,\infty)$, while the intermediate region of $(\delta-0.1,\delta+0.1)$ has been dealt with using an interpolation function.  Considering the numerical solution in the exterior region ($r>\delta+0.1$), we imposed the initial conditions: $f(\delta + 0.1) \approx 0.83$, $f'(\delta + 0.1) \approx 0.12$, 
$\Phi(\delta + 0.1) \approx 4.35$, 
$\Phi'(\delta + 0.1) = -0.39$ and $F(\delta + 0.1) \approx 0.999935$. 

In figure \ref{fig:1}, we depict the numerical solution for the Higgs function, the form of which is typical for global monopole models such as \cite{negative}. In figure \ref{fig:2}, we compare the numerically calculated metric function with the theoretical one, after the Israel matching (\ref{matchmetric}). These solutions are identical in the asymptotic regions, as it was expected, and they differ by up to $10^{-5}$ in the area around the monopole core. In figure \ref{fig:3} we present the numerical solution of the scalar field compared with the theoretical solution (\ref{scalar1}), which is valid in the exterior region ($r>\delta$). Interestingly, even close to the monopole core for $r>\delta$, the numerical solution of the scalar field, for reasonable initial conditions, is identical to the theoretical one. In all three figures we use three vertical lines denoting the thickness of the core region, which correspond to $r=\delta-0.1$, $r=\delta$ and $r=\delta+0.1$ respectively.   
\begin{figure}[H]
	\centering            
	\includegraphics[width=0.5\textwidth]{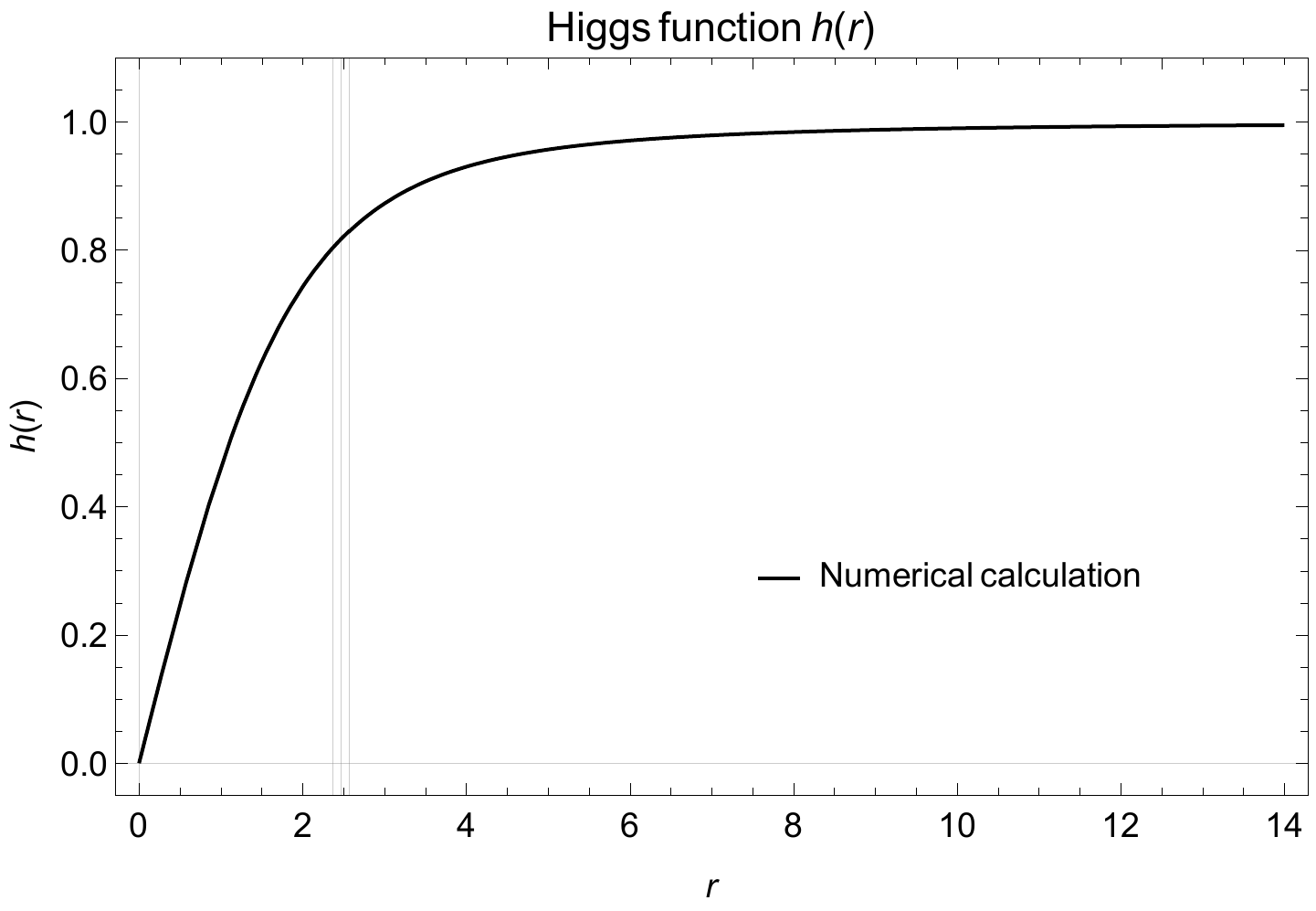}
	\caption{Numerical calculation of the Higgs function for $\eta=0.01$, $\xi=10^{4}$, $\alpha=-1.6\cdot 10^{-4}$, $\beta=8\cdot 10^{-9}$, $\lambda=-10^{-13}$ and $\kappa=1$. The three vertical lines denote the thickness of the core shell.}
	\label{fig:1}
\end{figure}

\begin{figure}[H]
    \centering            
    \includegraphics[width=0.5\textwidth]{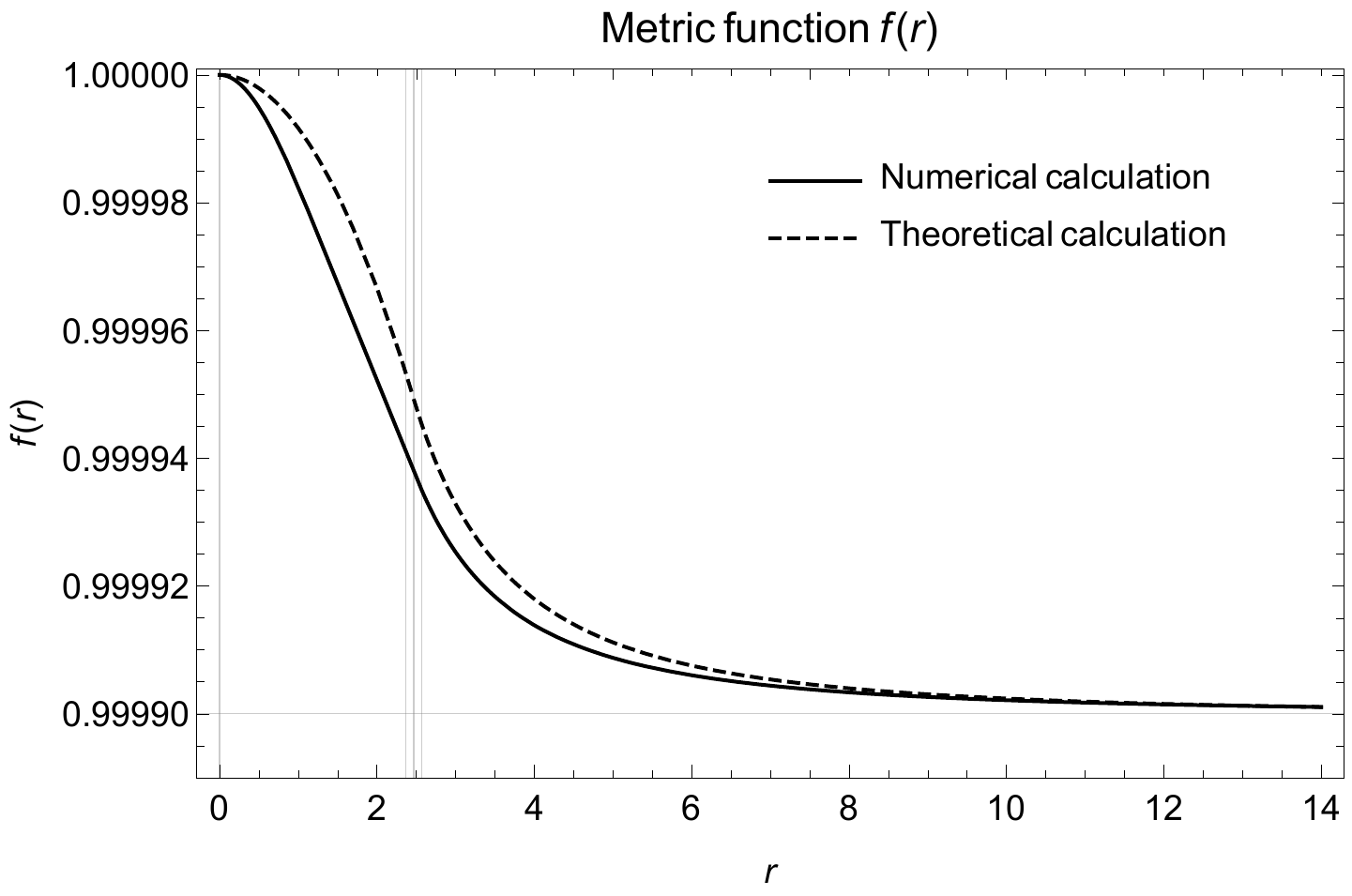}
    \caption{Numerical (solid line) and theoretical (dashed line) calculations of the metric function for $\eta=0.01$, $\xi=10^{4}$, $\alpha=-1.6\cdot 10^{-4}$, $\beta=8\cdot 10^{-9}$, $\lambda=-10^{-13}$ and $\kappa=1$. The three vertical lines denote the thickness of the core shell.}
    \label{fig:2}
\end{figure}

\begin{figure}[H]
    \centering            
    \includegraphics[width=0.5\textwidth]{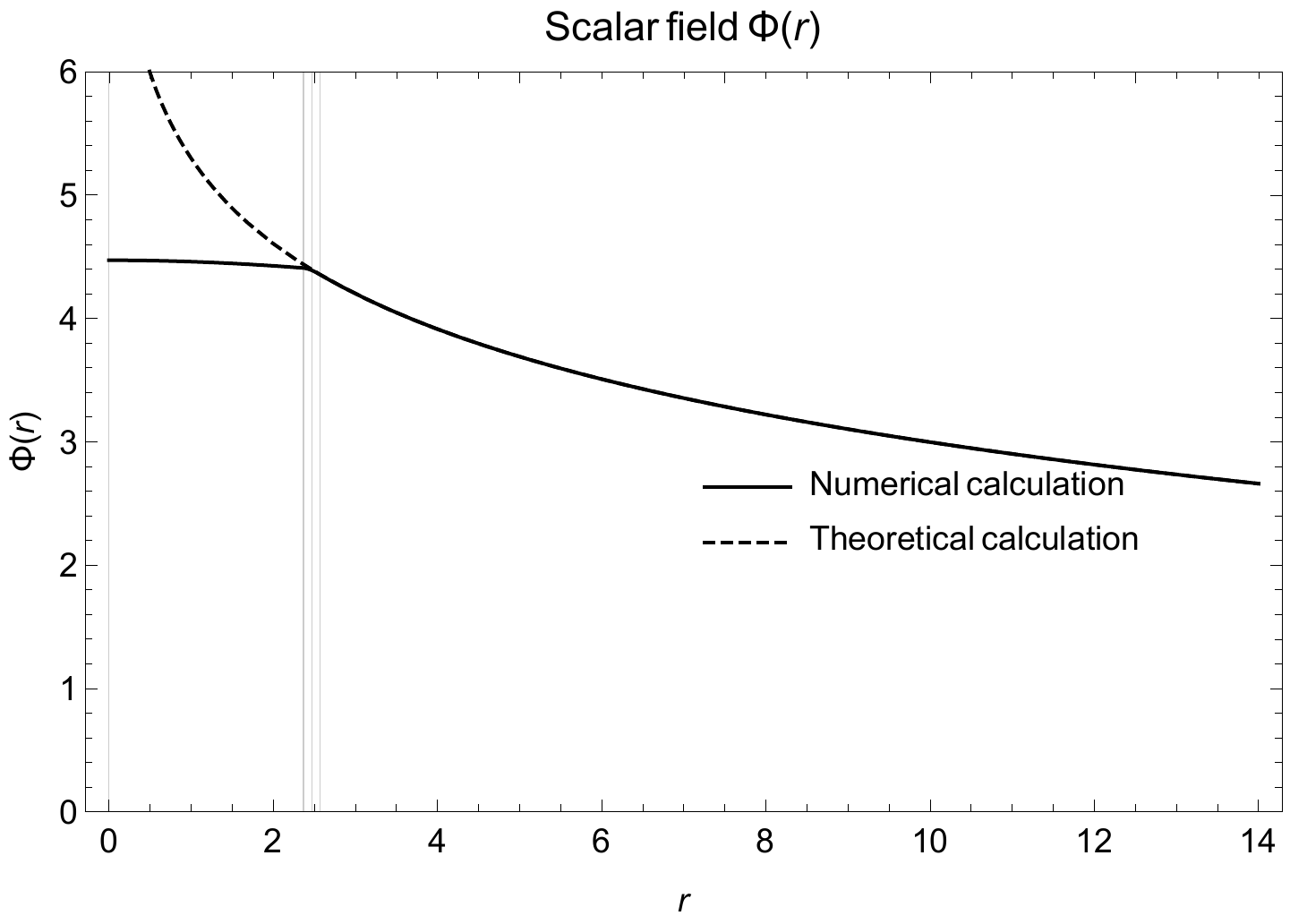}
    \caption{Numerical (solid line) and theoretical (dashed line) calculations of the Scalar field in the interior and exterior to the core regions of the global monopole in the eGB model, for the set of parameters given in fig.~\ref{fig:2}. The dilaton is approximately a step function near the origin. The three vertical lines denote the thickness of the core shell. The dashed line corresponds to the analytic scalar field solution of (\Ref{scalar1}), also valid for the exterior region $r>\delta+0.1$ of the monopole case.}
    \label{fig:3}
\end{figure}

\color{black} This completes our discussion on the numerical solution of the global monopole in the eGB model.
\color{black}

\section{Conclusions and Outlook: \color{black} potential phenomenology \color{black}}\label{sec:concl}

In this work, we have examined the existence of self-gravitating global monopole solutions of the type discussed in \cite{vilenkin},  but in the framework of the so-called extended Gauss-Bonnet (eGB) gravitational theory \eqref{action}, involving among other terms, the coupling of a scalar field $\Phi$ to the Gauss-Bonnet higher-curvature combination, with a coupling constant $\alpha$. As a consequence of this coupling, the embedding \eqref{scal} of the Higgs-triplet-$\chi$ sector to the eGB theory \eqref{action}, implies the possibility of regularising the singularity region $r \to 0$ at the centre of the monopole with a de-Sitter type spacetime, up to the position of the core $\delta$, which is then matched to an exterior to the monopole core  space-time \eqref{matchmetric}, which for $r \to \infty$ asymptotes ({\it cf.} \eqref{asym}) to a Minkowski space-time with a conical deficit  given by $ \eta^2 \, 8\pi \, G_{\rm N} $. One may then define a global metric \eqref{metricglobal} to cover the entire spacetime, including the interior of the monopole.

The existence of a conical deficit, as in the case of the (unstable) monopole of \cite{vilenkin}, would imply similar phenomenology as for the standard global monopole, discussed in section \ref{sec:intro}, albeit in our case the monopole has a positive mass, and thus it is likely to be stable. The positivity of the ADM mass of the monopole
is due to appropriate restrictions of the coupling constant $\alpha$, \eqref{rangealpha}, 
which when satisfied, guarantee the dynamical (through the appropriate Israel matching conditions) determination of a positive core radius $\delta$ and a positive ADM mass of the global monopole, \eqref{deltaM}. One verifies easily that in the limit $\alpha \to 0$, one obtains the standard negative-monopole mass result of \cite{negative}, \eqref{rangealpha}, for the monopole mass of \cite{vilenkin}. 

\color{black} Some remarks are in order here, regarding the cosmology, as well as galactic phenomenology of such objects. Actually, as already remarked in section \ref{sec:intro}, the phenomenology is rather generic to any object that leads to an asymptotic spacetime with a conical deficit. The global monopole discussed in this work, within the context of eGB gravity, is a local self-gravitating configuration involving scalar `hair' due to the non constant dilaton. The cosmology of the eGB model of \cite{Fernandes:2021dsb} {\it per se} is a topic by itself, which we would not like to enter here (for some preliminary discussion on such a topic in the absence of a deficit see~\cite{Fernandes:2021dsb,Fernandes:2022zrq}). Nonetheless, if the theory turns out to be physical and to contain such stable configurations, they could indeed play a r\^ole as cosmological dark matter candidates, since they are characterised by a positive ADM mass, provided, of course, that they exist in sufficiently large populations in the early Universe, and have masses in appropriate windows below the inflationary scale to avoid dilution by inflation, but also not to overclose the Universe. Irrespectively, though, of their potential cosmological importance, if localised populations of global monopoles exist in some regions of the early universe, then scattering of CMB photons on them will lead, as already mentioned, to Einstein rings~\cite{papavassiliou}. This is due to the fact that the scattering of electrically neutral massless particles on such space–times is described by amplitudes that exhibit resonant
behaviour when the forward-scattering and deficit angles coincide, leading to ring-like structures where the cross
sections are very large (formally divergent, appropriately regularised as we discuss below), a phenomenon termed “singular lensing” in the last reference of~\cite{papavassiliou}.

Specifically, as discussed in the last reference of \cite{papavassiliou}, for the physically interesting case of small deficits ($\eta^2 \, 8\pi\, G_{\rm N} \ll 1$, as expected in theories of phenomenological interest where the symmetry breaking mass scale $\eta$ is much smaller than the Planck scale) scattering of massless electrically neutral particles (such as CMB photons) of a global monopole background, will result in a differential cross section:
\bea\label{analyticdcs}
\frac{d \sigma }{d \Omega} & \stackrel{\theta \ge -\pi \tilde \zeta} =& \frac{1}{8\, \omega^2} \, \frac{{\rm sin}^2\pi \tilde \zeta}{\left({\rm cos}\pi \tilde \zeta - {\rm cos}\theta \right)^3} \, \left[ 1 - \frac{\pi\, (1-b^2)}{4\,b} \frac{\left({\rm cos}\pi \tilde \zeta - {\rm cos}\theta \right) }{{\rm sin}\pi \tilde \zeta}\, \right]^2  \nonumber \\
&=& \frac{1}{64\, \omega^2} \,  \frac{{\rm sin}^2\pi \tilde \zeta}{\left({\rm sin}(\frac{\Delta}{2}) \, {\rm sin}(\frac{\Delta}{2} + |\pi\tilde \zeta|) \right)^3} \, \left[1 - \frac{\pi\, (1-b^2)}{2\,b}\, 
\frac{\left({\rm sin}(\frac{\Delta}{2}) \, {\rm sin}(\frac{\Delta}{2} + |\pi \tilde \zeta|) \right)}{{\rm sin}\pi\tilde \zeta}\, \right]^2~,\nonumber \\
\eea
where we used the convenient notation of the last reference in \cite{papavassiliou}, with $\omega$ the energy of the incident particle/wave, $\theta$ the scattering angle, $b^2=1 - \eta^2 \, 8\pi \, G_{\rm N}  $ related to the conical deficit angle $2 \tilde \zeta \equiv 1 - b^2 = 8\pi\, \eta^2\, G_{\rm N}$,  and 
in the second line of \eqref{analyticdcs}, we have expressed the result in terms of the (non-negative) parameter $\Delta~\equiv \theta - |\pi\tilde \zeta| \ge 0$, using ${\rm cos}\pi\tilde \zeta - {\rm cos}\theta = 2\, \Big({\rm sin}(\frac{\Delta}{2}) \, {\rm sin}(\frac{\Delta}{2} + |\pi\tilde \zeta|) \Big)$. 
This allows for an easier visualisation of the physical effects of the limit $\theta \to |\pi \tilde \zeta | \ne 0$, \emph{i.e.} when $0 < \Delta \ll |\pi\tilde \zeta| $. Indeed, the leading behaviour of the differential cross section (\ref{analyticdcs}), as $\theta \to |\pi\tilde \zeta | \ne 0$ ($0 < \Delta \ll |\pi\tilde \zeta|$), is
\be\label{leadingdcs}
\frac{d \sigma }{d \Omega}  \stackrel{\theta \to |\pi\tilde \zeta | \ne 0} \simeq \frac{{\rm sin}^2\pi\tilde \zeta}{8\, \omega^2 \, \Big({\rm cos}\pi\tilde \zeta  - {\rm cos}\theta\Big)^3} 
 \stackrel{0< \Delta \ll |\pi\tilde \zeta|}\simeq \,  \frac{1}{64\, \omega^2 \, |{\rm sin}\pi\tilde \zeta |\, {\rm sin}^3(\frac{\Delta}{2})} ~.
\ee
which diverges for $\Delta \to 0$, giving rise to the aforementioned lensing phenomenon (formation of Einstein ring-like structures)~\cite{papavassiliou}. The phenomenon is independent of the spin of the scattered particle, and hence it applies equally well to photons.
It goes without saying that, in practice, this divergence will be regulated by the experimental/observational angular resolution $\theta_{\rm res}$, which imposes a natural cut-off $\Delta \ge \theta_{\rm res}$ in the above expressions. This would imply that 
the maximum value of the differential cross section attained will be $\frac{d\sigma}{d\Omega}|_{\rm max} = \frac{d\sigma (\Delta=\theta_{\rm res})}{d\Omega}$.\footnote{In fact, as discussed in the last reference of \cite{papavassiliou}, $\theta_{\rm res}$ acts as a regulator of the singular scattering amplitude and the total cross section computed using the optical theorem.} We also remark that  in the region of scattering angles 
$\pi\tilde \zeta \ll \Delta$ (for $\pi\tilde \zeta \ll 1$),  we obtain from (\ref{analyticdcs}) a {\it suppressed} differential cross section
\be\label{analyticdc2}
\frac{d \sigma }{d \Omega}  \stackrel{|\pi\tilde \zeta | \ll \Delta\, , \,|\pi\tilde \zeta | \ll 1}\simeq \frac{(\pi\tilde \zeta)^2}{64\, \omega^2\,
  {\rm sin}^6(\frac{\Delta}{2})} \left[1 + {\rm sin}^2\left(\frac{\Delta}{2}\right) \right]^2~,
\ee
where we employed the approximation $\pi (1-b^2)/{2 b} \simeq - \pi\tilde \zeta >0$, for $|\pi \tilde \zeta | \ll 1$. 

In principle, therefore, by measuring the size of the ring-like region implied by \eqref{leadingdcs},  we can determine the parameter $b$, or, equivalently, the deficit angle in the asymptotic
spacetime metric \eqref{asymptflat}, and, therefore, the symmetry breaking scale $\eta$.
The lensing phenomenon is generic to such asymptotic spacetimes with a deficit and is independent of the 
underlying microscopic geometry which asymptotes to \eqref{asymptflat}. Hence, it is {\it not} particular to the global monopole. In the standard BV monopole~\cite{vilenkin} the deficit would be proportional to the mass of the monopole, and thus measuring $\eta$ would determine this mass. However, in the eGB case, the (positive) monopole ADM mass \eqref{deltaM} is a function not only of $\eta$ but also of the GB parameter $\alpha$ and the scalar self-interaction coupling $\xi$, hence by measuring the $\eta$ alone through the singular lensing phenomenon is not sufficient to infer an information on the global monopole mass. One needs to determine also these other two parameters $\alpha, \xi$ independently. A combination of these parameters enter also the monopole ``core'' $\delta$ \eqref{deltaM} (where the matching of the de Sitter interior  to the exterior spacetime has been performed). Ordinary gravitational lensing of photons by such compact objects (associated with scattering angles outside the singular lensing regime), which in general depends on both the mass $M$ and the core size $\delta$ of the global monopoles, will provide additional constraints on the three parameters $\eta, \alpha$ and $\xi$ of the eGB theory of the global monopole.  The fact that the eGB theory also admits black hole solutions~\cite{Fernandes:2021dsb}, and in general exhibits a rich phenomenology~\cite{Fernandes:2022zrq}, which allows us to constrain independently the GB parameter $\alpha$, features that are expected to be valid also in the presence of a deficit angle (albeit with non asymptotically flat solution ({\it cf.} \eqref{scalardef}), implies that one has at their disposal sufficient phenomenological and observational tools to constrain observationally/experimentally all three parameters $\eta$, $\xi$ and $\alpha$. In this way we can  place constraints not only on the mass and core radius of such (stable) eGB global monopoles, but also on their local abundance ({\it e.g.} within our galaxy).  We remark that, in view of the local character of the solution, the eGB global monopoles, if stable, could play the r\^ole of ``galactic dark matter'' candidates, even if they turn out not to be the dominant cosmological dark matter. We hope therefore the current work will simulate such searches  in the future.
\color{black}

The positivity of the ADM mass of the global monopole is perhaps an encouraging factor towards its stability. Nonetheless, in the present article we do not want to make strong claims until a detailed stability analysis on the solution is performed, along the corresponding studies in black hole physics~\cite{stab}. It is only when such a stability analysis is complete, that we can enter with a definite answer the debate on the stability of the global monopole in eGB gravity. 
\color{black} To verify that our configuration is stable, a detailed time-dependent perturbation analysis on the metric configurations needs to be implemented, which, as we have already mentioned, is out of the scope of this paper, and constitutes a task for a future work. We do note at this stage that gravitational wave analysis on a global monopole spacetime, required for the stability study, is a highly non-trivial subject and it would be a stand-alone work in this case due to the effect of the solid deficit angle on the boundary conditions of the evolution of the perturbations. Indeed, the presence of a deficit angle will probably imply, in view of the aforementioned singular lensing phenomenon \eqref{leadingdcs}, which would also characterise gravitational wave scattering off global monopoles, that linear stability studies as in the standard string-inspired GB black hole of \cite{stab} is not sufficient to demonstrate stability, and hence one would require the machinery of the full numerical relativity, which is out of the scope of this work.\footnote{We remark at  this point that, in the standard eGB theory of \cite{Fernandes:2021dsb} without symmetry breaking, and thus deficit angles, the black hole has been claimed to be a linearly stable configuration in \cite{Fernandes:2021ysi}, by arguing that, due to its {\it asymptotic flatness}, any time dependent linear perturbation of the black-hole solution, vanishes. However, such claims have been recently disputed in \cite{Tsujikawa:2022lww}, where it was argued that instabilities are present for such black holes in the even-parity angular linear perturbations both at the horison and at infinity. The present global monopole solution corresponds to a spacetime, which as we have argued before, is not asymptotically flat ({\it cf.} \eqref{asymptflat}, \eqref{scalardef}), due to the spacetime deficit angle, associated with the symmetry breaking scale $\eta$ of the global monopole. This is an important difference which affects linear stability, in the way mentioned above.} The main subject of the paper was to show that regularized global monopoles with positive mass exist in the eGB gravity, which, as we have hopefully communicated to the reader, was itself not a trivial issue, given that it is a result which is absent in GR, and, in addition, remedies a main pathology of the global monopole in GR~\cite{vilenkin,negative}. \color{black}

Another issue we would like to look at is whether there is the possibility of inducing a magnetic monopole from the global monopole solution, following an analysis inspired by the work of \cite{sarkarNEM}. Perhaps, one way to do this is to allow for a Maxwell term in the action \eqref{scal}, with an appropriate form factor $-\frac{1}{4} \, f(\Phi) F_{\mu\nu}F^{\mu\nu}$, such that for constant scalar field  $\Phi=\Phi_0$, this factor goes to 1, $f(\Phi \to \Phi_0) =1$, so that one recovers the standard electromagnetic interactions in the presence of gravity. It would be interesting to see whether, under appropriate selection of the function $f(\Phi)$, one may obtain magnetic field configurations of magnetic monopole type, as in \cite{sarkarNEM}. Such an analysis is postponed for a future work. {\it affaire \`a suivre ...}

\section*{Acknowledgments}

The work of N.C. is supported by the research project
of the National Technical University of Athens (NTUA) 65232600-ACT-MTG: {\it Alleviating Cosmological Tensions
Through Modified Theories of Gravity}; that of N.E.M. is supported in part by the UK Science and Technology Facilities research Council (STFC) under the research grant ST/T000759/1; and the work of D.T. is supported by a National Technical University of Athens Master-programme ({\it Physics and Technological Applications}) award of scientific excellence.
N.E.M.  also acknowledges participation in the COST Association Action CA18108 ``{\it Quantum Gravity Phenomenology in the Multimessenger Approach (QG-MM)}''.


\begin{thebibliography}{99}

\bibitem{vilenkin} M.~Barriola and A.~Vilenkin,
  ``Gravitational Field of a Global Monopole,''
  Phys.\ Rev.\ Lett.\  {\bf 63}, 341 (1989).
  doi:10.1103/PhysRevLett.63.341
  %%CITATION = doi:10.1103/PhysRevLett.63.341;%%
  %450 citations counted in INSPIRE as of 17 Jun 2016


\bibitem{papavassiliou} 
  P.~O.~Mazur and J.~Papavassiliou,
  ``Gravitational scattering on a global monopole,''
  Phys.\ Rev.\ D {\bf 44}, 1317 (1991).
  doi:10.1103/PhysRevD.44.1317;
  %%CITATION = doi:10.1103/PhysRevD.44.1317;%%
  %12 citations counted in INSPIRE as of 17 Feb 2017
  H.~Ren,
  ``Fermions in a global monopole background,''
  Phys.\ Lett.\ B {\bf 325}, 149 (1994)
  doi:10.1016/0370-2693(94)90085-X
  [hep-th/9312074];
  %%CITATION = doi:10.1016/0370-2693(94)90085-X;%%
  %6 citations counted in INSPIRE as of 17 Feb 2017
E.~R.~Bezerra de Mello and C.~Furtado,
  ``The Nonrelativistic scattering problem by a global monopole,''
  Phys.\ Rev.\ D {\bf 56}, 1345 (1997).
  doi:10.1103/PhysRevD.56.1345
  %%CITATION = doi:10.1103/PhysRevD.56.1345;%%
  %29 citations counted in INSPIRE as of 17 Feb 2017
A.~A.~Roderigues Sobreira and E.~R.~Bezerra de Mello,
  ``The Classical and quantum analysis of a charged particle on the space-time produced by a global monopole,''
  Grav.\ Cosmol.\  {\bf 5}, 177 (1999)
  [hep-th/9809212];
  %%CITATION = HEP-TH/9809212;%%
  %5 citations counted in INSPIRE as of 17 Feb 2017
N.~E.~Mavromatos and J.~Papavassiliou,
``Singular lensing from the scattering on special space\textendash{}time defects,''
Eur. Phys. J. C \textbf{78}, no.1, 68 (2018)
doi:10.1140/epjc/s10052-018-5542-5
[arXiv:1712.03395 [hep-ph]].
%11 citations counted in INSPIRE as of 18 Dec 2022

\bibitem{debate} A.~S.~Goldhaber,
  ``Collapse of a 'Global Monopole.',''
  Phys.\ Rev.\ Lett.\  {\bf 63}, 2158 (1989).
  doi:10.1103/PhysRevLett.63.2158;
  %%CITATION = doi:10.1103/PhysRevLett.63.2158;%%
  %25 citations counted in INSPIRE as of 09 Oct 2016
In the original suggestion of Goldhaber that global monopoles are not stable against ``angular'' collapse, there is an ongoing debate on this issue; for a partial list of references  see:
S.~H.~Rhie and D.~P.~Bennett,
  ``Global monopoles do not 'collapse',''
  Phys.\ Rev.\ Lett.\  {\bf 67}, 1173 (1991).
  doi:10.1103/PhysRevLett.67.1173;
  %%CITATION = doi:10.1103/PhysRevLett.67.1173;%%
  %8 citations counted in INSPIRE as of 09 Oct 2016
L.~Perivolaropoulos,
  ``Instabilities and interactions of global topological defects,''
  Nucl.\ Phys.\ B {\bf 375}, 665 (1992).
  doi:10.1016/0550-3213(92)90115-R;
  %%CITATION = doi:10.1016/0550-3213(92)90115-R;%%
  %27 citations counted in INSPIRE as of 09 Oct 2016
G.~W.~Gibbons, M.~E.~Ortiz, F.~Ruiz Ruiz and T.~M.~Samols,
  ``Semilocal strings and monopoles,''
  Nucl.\ Phys.\ B {\bf 385}, 127 (1992)
  doi:10.1016/0550-3213(92)90097-U
  [hep-th/9203023];
  %%CITATION = doi:10.1016/0550-3213(92)90097-U;%%
  %67 citations counted in INSPIRE as of 09 Oct 2016
M.~Hindmarsh,
  ``Semilocal topological defects,''
  Nucl.\ Phys.\ B {\bf 392}, 461 (1993)
  doi:10.1016/0550-3213(93)90681-E
  [hep-ph/9206229];
  %%CITATION = doi:10.1016/0550-3213(93)90681-E;%%
  %80 citations counted in INSPIRE as of 09 Oct 2016
G.~Arreaga, I.~Cho and J.~Guven,
  ``Stability of selfgravitating magnetic monopoles,''
  Phys.\ Rev.\ D {\bf 62}, 043520 (2000)
  doi:10.1103/PhysRevD.62.043520
  [gr-qc/0001078];
  %%CITATION = doi:10.1103/PhysRevD.62.043520;%%
  %8 citations counted in INSPIRE as of 09 Oct 2016
A.~Achucarro and J.~Urrestilla,
  ``The (In)stability of global monopoles revisited,''
  Phys.\ Rev.\ Lett.\  {\bf 85}, 3091 (2000)
  doi:10.1103/PhysRevLett.85.3091
  [hep-ph/0003145];
  %%CITATION = doi:10.1103/PhysRevLett.85.3091;%%
  %16 citations counted in INSPIRE as of 09 Oct 2016
  R.~Gregory and C.~Santos,
  ``Space-time structure of the global vortex,''
  Class.\ Quant.\ Grav.\  {\bf 20}, 21 (2003)
  doi:10.1088/0264-9381/20/1/302
  [hep-th/0208037];
  %%CITATION = doi:10.1088/0264-9381/20/1/302;%%
  %16 citations counted in INSPIRE as of 09 Oct 2016
  E.~R.~Bezerra de Mello,
  ``Reply on comment on `Gravitating magnetic monopole in the global monopole space-time',''
  Phys.\ Rev.\ D {\bf 68}, 088702 (2003)
  doi:10.1103/PhysRevD.68.088702
  [hep-th/0304029];
  %%CITATION = doi:10.1103/PhysRevD.68.088702;%%
  %7 citations counted in INSPIRE as of 24 Feb 2017  
  S.~B.~Gudnason and J.~Evslin,
  ``Global monopoles of charge 2,''
  Phys.\ Rev.\ D {\bf 92}, no. 4, 045044 (2015)
  doi:10.1103/PhysRevD.92.045044
  [arXiv:1507.03400 [hep-th]].
  %%CITATION = doi:10.1103/PhysRevD.92.045044;%%
  
  \bibitem{negative} D.~Harari and C.~Lousto,
  ``Repulsive gravitational effects of global monopoles,''
  Phys.\ Rev.\ D {\bf 42}, 2626 (1990).
  doi:10.1103/PhysRevD.42.2626
  %%CITATION = doi:10.1103/PhysRevD.42.2626;%%
  %121 citations counted in INSPIRE as of 24 Jun 2016

\bibitem{bronnikov} K.~A.~Bronnikov, B.~E.~Meierovich and E.~R.~Podolyak,
  ``Global monopole in general relativity,''
  J.\ Exp.\ Theor.\ Phys.\  {\bf 95}, 392 (2002)
  [Zh.\ Eksp.\ Teor.\ Fiz.\  {\bf 122}, 459 (2002)]
  doi:10.1134/1.1513811
  [gr-qc/0212091].
  %%CITATION = doi:10.1134/1.1513811;%%
  %10 citations counted in INSPIRE as of 24 Feb 2017


\bibitem{sarkarNEM} N.~E.~Mavromatos and S.~Sarkar,
``Magnetic monopoles from global monopoles in the presence of a Kalb-Ramond Field,''
Phys. Rev. D \textbf{95}, no.10, 104025 (2017)
doi:10.1103/PhysRevD.95.104025
[arXiv:1607.01315 [hep-th]];
%29 citations counted in INSPIRE as of 18 Dec 2022
``Regularized Kalb-Ramond magnetic monopole with finite energy,''
Phys. Rev. D \textbf{97}, no.12, 125010 (2018)
doi:10.1103/PhysRevD.97.125010
[arXiv:1804.01702 [hep-th]].
%15 citations counted in INSPIRE as of 18 Dec 2022

\bibitem{shnir} Y.~M.~Shnir,
``Magnetic Monopoles,''
Springer, 2005,
ISBN 978-3-540-25277-1, 978-3-540-29082-7
doi:10.1007/3-540-29082-6
%35 citations counted in INSPIRE as of 18 Dec 2022

\bibitem{mitsou} For a review of recent developments see: %\cite{Mavromatos:2020gwk}
N.~E.~Mavromatos and V.~A.~Mitsou,
``Magnetic monopoles revisited: Models and searches at colliders and in the Cosmos,''
Int. J. Mod. Phys. A \textbf{35}, no.23, 2030012 (2020)
doi:10.1142/S0217751X20300124
[arXiv:2005.05100 [hep-ph]], and references therein.
%47 citations counted in INSPIRE as of 18 Dec 2022

	
	%\cite{Fernandes:2022zrq}
	
	
		
	\bibitem{Fernandes:2021dsb}
	P.~G.~S.~Fernandes,
	``Gravity with a generalized conformal scalar field: theory and solutions,''
	Phys. Rev. D \textbf{103}, no.10, 104065 (2021)
	doi:10.1103/PhysRevD.103.104065
	[arXiv:2105.04687 [gr-qc]].
	%\cite{Lu:2020iav}
	
		%\cite{Bocharova:1970skc}
	
		
	
	\bibitem{Bocharova:1970skc}
	N.~M.~Bocharova, K.~A.~Bronnikov and V.~N.~Melnikov,
	Vestn. Mosk. Univ. Ser. III Fiz. Astron. 6, 706-709 (1970) (in Russian).
	%1 citations counted in INSPIRE as of 14 Dec 2022
	
	
	%\cite{Bekenstein:1974sf}
	\bibitem{Bekenstein:1974sf}
	J.~D.~Bekenstein,
	``Exact solutions of Einstein conformal scalar equations,''
	Annals Phys. \textbf{82}, 535-547 (1974)
	doi:10.1016/0003-4916(74)90124-9
	%419 citations counted in INSPIRE as of 14 Dec 2022
		
		
		
	\bibitem{Lu:2020iav}
	H.~Lu and Y.~Pang,
	``Horndeski gravity as $D \rightarrow 4$ limit of Gauss-Bonnet,''
	Phys. Lett. B \textbf{809}, 135717 (2020)
	doi:10.1016/j.physletb.2020.135717
	[arXiv:2003.11552 [gr-qc]].
	%177 citations counted in INSPIRE as of 14 Dec 2022
	%18 citations counted in INSPIRE as of 14 Dec 2022
	%\cite{Barriola:1989hx}
	
	
		
	\bibitem{Fernandes:2022zrq}
	P.~G.~S.~Fernandes, P.~Carrilho, T.~Clifton and D.~J.~Mulryne,
	``The 4D Einstein\textendash{}Gauss\textendash{}Bonnet theory of gravity: a review,''
	Class. Quant. Grav. \textbf{39}, no.6, 063001 (2022)
	doi:10.1088/1361-6382/ac500a
	[arXiv:2202.13908 [gr-qc]].
	%28 citations counted in INSPIRE as of 14 Dec 2022
	%\cite{Fernandes:2021dsb}

	
		
	%\cite{Babichev:2022awg}
	\bibitem{Babichev:2022awg}
	E.~Babichev, C.~Charmousis, M.~Hassaine and N.~Lecoeur,
	``Conformally coupled theories and their deformed compact objects: From black holes, radiating spacetimes to eternal wormholes,''
	Phys. Rev. D \textbf{106}, no.6, 064039 (2022)
	doi:10.1103/PhysRevD.106.064039
	[arXiv:2206.11013 [gr-qc]].
	%2 citations counted in INSPIRE as of 14 Dec 2022
	
	

%\cite{Kanti:1997br}
\bibitem{stab} see, for instance:
P.~Kanti, N.~E.~Mavromatos, J.~Rizos, K.~Tamvakis and E.~Winstanley,
``Dilatonic black holes in higher curvature string gravity. 2: Linear stability,''
Phys. Rev. D \textbf{57}, 6255-6264 (1998)
doi:10.1103/PhysRevD.57.6255
[arXiv:hep-th/9703192 [hep-th]];
%120 citations counted in INSPIRE as of 18 Dec 2022
%\cite{Mavromatos:1995kc}
N.~E.~Mavromatos and E.~Winstanley,
``Aspects of hairy black holes in spontaneously broken Einstein Yang-Mills systems: Stability analysis and entropy considerations,''
Phys. Rev. D \textbf{53}, 3190-3214 (1996)
doi:10.1103/PhysRevD.53.3190
[arXiv:hep-th/9510007 [hep-th]];
%60 citations counted in INSPIRE as of 18 Dec 2022	
%\cite{Winstanley:1995iq}
E.~Winstanley and N.~E.~Mavromatos,
``Instability of hairy black holes in spontaneously broken Einstein Yang-Mills Higgs systems,''
Phys. Lett. B \textbf{352}, 242-246 (1995)
doi:10.1016/0370-2693(95)00562-Y
[arXiv:hep-th/9503034 [hep-th]].
%26 citations counted in INSPIRE as of 18 Dec 2022	
%\cite{Hertog:2004bb}
T.~Hertog and K.~Maeda,
``Stability and thermodynamics of AdS black holes with scalar hair,''
Phys. Rev. D \textbf{71}, 024001 (2005)
doi:10.1103/PhysRevD.71.024001
[arXiv:hep-th/0409314 [hep-th]].
%66 citations counted in INSPIRE as of 18 Dec 2022
%\cite{Tamaki:2003ah}
T.~Tamaki, T.~Torii and K.~i.~Maeda,
``Stability analysis of black holes via a catastrophe theory and black hole thermodynamics in generalized theories of gravity,''
Phys. Rev. D \textbf{68}, 024028 (2003)
doi:10.1103/PhysRevD.68.024028
%13 citations counted in INSPIRE as of 18 Dec 2022
%\cite{Torii:1998gm}
T.~Torii and K.~i.~Maeda,
``Stability of a dilatonic black hole with a Gauss-Bonnet term,''
Phys. Rev. D \textbf{58}, 084004 (1998)
doi:10.1103/PhysRevD.58.084004
%39 citations counted in INSPIRE as of 18 Dec 2022

%\cite{Fernandes:2021ysi}
\bibitem{Fernandes:2021ysi}
P.~G.~S.~Fernandes, P.~Carrilho, T.~Clifton and D.~J.~Mulryne,
``Black holes in the scalar-tensor formulation of 4D Einstein-Gauss-Bonnet gravity: Uniqueness of solutions, and a new candidate for dark matter,''
Phys. Rev. D \textbf{104}, no.4, 044029 (2021)
doi:10.1103/PhysRevD.104.044029
[arXiv:2107.00046 [gr-qc]].
%15 citations counted in INSPIRE as of 02 Feb 2023	

\bibitem{Tsujikawa:2022lww}
S.~Tsujikawa,
``Instability of hairy black holes in regularized 4-dimensional Einstein-Gauss-Bonnet gravity,''
Phys. Lett. B \textbf{833}, 137329 (2022)
doi:10.1016/j.physletb.2022.137329
[arXiv:2205.09932 [gr-qc]].
%3 citations counted in INSPIRE as of 02 Feb 2023


\end{thebibliography}
\end{document}